\documentclass[twocolumn,showpacs,unsortedaddress,prl]{revtex4}
\usepackage{graphicx}
\usepackage{hyperref}

\def\gz{\ifmmode{Z\hskip -4.8pt Z}
    \else{\hbox{$Z\hskip -4.8pt Z$}}\fi}

\newcommand{\be}{\begin{equation}}
\newcommand{\ee}{\end{equation}}
\newcommand{\bea}{\begin{eqnarray}}
\newcommand{\eea}{\end{eqnarray}}

\begin{document}

\title{Dynamical mean field theory of an effective three-band model for Na$_x$CoO$_2$}
\author{A. Bourgeois$^1$, A.~A.~Aligia$^2$, M.~J.~ Rozenberg$^1$}
\affiliation{$^1$Laboratoire de Physique des Solides, Univ. Paris-Sud,
CNRS UMR-8502, 91405 Orsay cedex, France} 
\affiliation{$^2$Centro
At\'omico Bariloche and Instituto Balseiro, Comisi\'on Nacional de
Energ\'{i}a At\'omica, 8400 Bariloche, Argentina}
\date{\today}

\begin{abstract}
We derive an effective Hamiltonian for highly correlated $t_{2g}$
states centered at the Co sites of Na$_x$CoO$_2$. The essential
ingredients of the model are an O mediated hopping, a trigonal
crystal-field splitting, and on-site effective interactions
derived from the exact solution of a multi-orbital 
model in a CoO$_6$
cluster, with parameters determined previously.
The effective model is solved by
dynamical mean-field theory (DMFT). We obtain a Fermi surface (FS) and
electronic dispersion that agrees well with angle-resolved
photoemission spectra (ARPES). 
Our results also elucidate the origin of the "sinking-pockets"
in different doping regimes.
\end{abstract}


\maketitle

The construction of the appropriate low-energy Hamiltonian to
describe a highly correlated system is a crucial task for an
advance in its physical understanding. A clear example is the
case of the superconducting cuprates. The starting point for the
description of those materials is a three-band model containing the
most relevant Cu and O orbitals. The parameters of that model were
determined by constrained-density-functional theory \cite{cdf}. On
the basis of the exact solution of the multi-band model in a
CuO$_4$ cluster (containing one Cu atom and its four nearest
neighbors), Zhang and Rice suggested that the essential low-energy
physics of the model is captured by a one-band model containing
only effective Cu orbitals \cite{zr}. 
This has been confirmed by
systematic derivations of the ensuing one-band Hubbard and $t-J$
models \cite{hub,pt}. These models have led to a considerable
progress in the understanding of the high-$T_c$ cuprates. Similar
low-energy effective models were derived and used successfully to
explain the properties of nickelates \cite{oles,epl} and other
transition metal oxides \cite{feiner}.

In the cobaltates Na$_x$CoO$_2$ 
a consensus has not yet been reached on the appropriate low-energy effective Hamiltonian, 
as different approaches have provided conflicting
results. 
The cobaltates present a clear cut example of strong correlation
effects. Not only by its rich phase diagram that includes a charge
ordered insulator and a superconducting state at intermediate dopings,
but also by the complete failure of the ab-initio band structure calculations
to describe the shape and topology of the Fermi surface measured in ARPES
experiments \cite{arpes1,arpes2}. Specifically, first-principles
calculations done in the local-density approximation (LDA) \cite{lda}
predicted a Fermi surface with six prominent hole pockets along the
$\Gamma - K$ direction, which were never detected in photoemission.
In addition, the ARPES experiments have revealed the presence of
dispersive features at the momenta positions where those pockets were expected 
but they were observed at about 0.2~eV beneath the Fermi surface. Thus, they were
termed "sinking pockets" and are still awaiting a clear physical interpretation.  

Initial theoretical progress was seemingly achieved
by Zhou {\it et al.} \cite{zhou} who included correlation effects
on top of a tight-binding model fit to the band-structure from 
first principles calculations in the
local-density approximation
LDA \cite{lda}. They showed that correlation
effects may in fact wipe out the pockets by a reduction of the bandwidth
of the bands crossing the Fermi energy. However, the approach of Zhou {\it et al.} 
relied on a simplified static 
Gutzwiller approximation (GA) where the rather 
unrealistic assumption of an infinite strength for
the local effective $t_{2g}$ Coulomb repulsion $U$ is made.
In a different approach to the problem, Ishida {\it et al.} \cite{ishida-liebsch}
used the more elaborate DMFT methodology to treat the correlation effects, on
top of a similar LDA-derived tight-binding Hamiltonian. Significantly, the DMFT 
method allowed for the assumption of 
finite effective Coulomb interactions. 
The main finding of that work was the 
prediction that the effect of $U$ is, 
in marked contrast to the GA,
to actually {\em increase} the size of the LDA pockets, that get stabilized 
due to charge transfer between the $e'_g$ 
and $a_{1g}$ bands. 

Marianetti {\it et al.} \cite{marianetti} using
a DMFT calculation similar to Ishida {\it et al.} found that the pockets
can be made to disappear for sufficiently large values of $U$ (above 6 eV), 
which explained their absence in the 
infinite $U$ calculation. 
Although for realistic values of $U$ the pockets still remained, those authors
also pointed out that using the $e'_g - a_{1g}$ crystal field splitting
as a free fitting parameter, they could eventually be made to disappear.
More recently, Liebsch and Ishida \cite{liebsch-ishida}, critically discussed 
the various previous approaches that were based on the LDA band-structure as the starting point
for the calculation of correlation effects. They concluded that, at
values $U \sim 3$ eV which they considered realistic, the presence of 
pockets in the Fermi surface is always predicted. 
They argued that this feature, which is in conflict with ARPES data, is robust with respect to
the details of the LDA-fits and to the form of the interaction term.

From a more general perspective, one may expect that a simple-minded identification 
of the LDA-derived conduction bands as
the relevant manifold where correlations are to be included through
Hubbard-like interactions in an LDA+DMFT treatment 
may not be fully justified 
when systems have a strong covalent character, as it is the case of the cobaltates.
In particular it was shown that the above
procedure fails in NiO, and agreement with experiments in LDA+DMFT calculations
is only achieved once the O bands are explicitly included in the model \cite{vol}.
Interestingly, the results of that approach also agree 
with
results from effective models where the O atoms have been integrated out
using low-energy reduction procedures that take into account
correlations from the beginning \cite{vol,oles}.

We propose to address the problem
of the low-energy description of the band-structure of the cobaltates
by taking a different approach and altogether leave the LDA as the 
starting point of our calculation. 
Thus, in this Letter we perform a low-energy reduction to derive an effective
Hamiltonian $H_{\rm eff}$, that includes the determination of the values of effective local
repulsive interactions, and then study its physical behaviour using DMFT. 
The derivation of $H_{\rm eff}$ follows the ideas of previous research in the cuprates which used
the cell-perturbation method \cite{hub} and non-orthogonal
Zhang-Rice singlets \cite{zr,pt} constructions. Basically, the procedure is to divide
the system in different cells that are solved exactly, and retain their lowest
energy states. Then, one includes the inter-cell terms along with the effect of
the other states as perturbations to this low-energy subspace. The
resulting effective Hamiltonian  differs substantially
from those previously adopted. In particular, our calculated value of
$U$ is significantly smaller.
again raising questions on the justification of the assumption of an infinite value for 
the Coulomb interaction made in Gutzwiller-type of approaches. This observation 
also applies to a recent Gutzwiller Density-Functional calculation that reports good agreement 
with ARPES data, in which $U \gtrsim 3-5$ eV was assumed \cite{gut}.
A previously derived $H_{\rm eff}$ also assumed infinite effective on-site
$t_{2g}$ Coulomb repulsions instead of calculating them \cite{bourgeois}.

We start from the exact solution of a CoO$_6$ 
cluster model containing all 3d
orbitals of a Co atom and all 2p orbitals of its six
nearest-neighbor O atoms, assuming cubic (O$_h$) symmetry and
neglecting spin-orbit coupling. All interactions inside the 3d
shell are included \cite{kroll}. The parameters were determined
fitting XAS experiments and its polarization dependence
\cite{kroll}. The results, which agree with previous similar
studies \cite{wu}, show a large Co-O covalency and an intra-orbital
repulsion $U_m=4.5$ eV, larger than the Co-O charge-transfer energy.
The subscript $m$ refers to the original
multiband model, to distinguish $U_m$ from the corresponding
repulsion $U$ of the effective model, which as shown below, is
strongly reduced due to Co-O covalency. 
We recall
that in the cuprates, $U_m \sim 10$ eV \cite{cdf},
while in their effective low-energy one-band Hubbard model $U \sim
3$ eV \cite{hub}.

The effective model $H_{\rm eff}$ is obtained mapping the ground state
of the CoO$_6$ cluster with four holes 
onto the on-site vacuum of $H_{\rm eff}$ (no
$t_{2g}$ holes at a Co site, i.e. Co$^{+3}$), and the 6-fold
degenerate (spin doublet and orbital triplet) ground state for
five holes onto the corresponding states with one $t_{2g}$ hole of
$H_{\rm eff}$. Details of the mapping are given in Ref.
\cite{bourgeois}. 
We remark that Co$^{+3}$ and Co$^{+4}$ in $H_{\rm eff}$ actually represent
highly correlated states with a Co valence near 2.04 and 2.56 respectively \cite{kroll}. 
$H_{\rm eff}$ reads

\be
H_{\rm eff}=H_0+H_I,
\label{h}
\ee
where the ``non-interacting" part can be written as

\be
H_0= 
\sum_{i,j} \sum_{\alpha,\alpha',\sigma} \bigl( t^{ij}_{\alpha\alpha'} 
+ t'^{\,ij}_{\alpha}\delta_{\alpha\alpha'}
+ D^{\vphantom{i}}_{\alpha\alpha'} \delta_{ij} \bigr) 
d^{\dag}_{i\alpha\sigma} d_{j\alpha'\sigma},
\label{h0}
\ee
where 
$d_{i\alpha \sigma }^{\dagger }$ creates a hole in the
$t_{2g}$ orbital $\alpha $ ($xy$, $yz$ or $zx$) with spin $\sigma$ 
at site $i$. However, physically this operator represents a non
trivial excitation of the same symmetry, which involves also 3d
$e_g$ orbitals of Co and 2p orbitals of nearest-neighbor O
sites.
$t'$ and $t$ correspond to the direct Co-Co hopping and to that mediated by O 2p$_\pi$ orbitals 
\cite{koshi} 
respectively. The latter is the most important one and has been calculated before using many-body eigenstates of the CoO$_6$ cluster \cite{bourgeois}.
Finally $D$ accounts for the trigonal crystal-field splitting $\Delta=3D$ between $e'_g$ and $a_{1g}$ orbitals. We take it from 
quantum-chemistry configuration-interaction calculations \cite{ll}. These
are the most reliable methods to determine crystal-field
excitations. Incidentally, it is known that while the LDA may provide a good
description of the ground state, it does not get the energy of excited states right.
Therefore it is not expected to provide accurate values for $D$ in a
highly correlated system.
Note that although $H_0$ has the form of a non-interacting Hamiltonian, the derivation of its parameters already 
involve many-body calculations \cite{bourgeois,ll}.
In fact, similarly as in the studies of cuprates, most of the original Co on-site interaction  
is already included in the derivation of
$H_{\rm eff}$ through the diagonalization of the
CoO$_6$ cluster.

The interacting part of $H_{\rm eff}$  is

\bea
H_{I}^{i} = \sum_{i} H_{I}^{i};~~~~~~~~~~~ H_{I}^{i} = U \sum_{\alpha } n_{i\alpha \uparrow } n_{i\alpha
\downarrow } + \nonumber \\
+ \frac{1}{2} \sum_{\alpha \neq \beta, \sigma \sigma
^{\prime }} (U^{\prime } n_{i\alpha \sigma} n_{i\beta
\sigma^{\prime }} + J d_{i\alpha \sigma }^{\dagger }d_{i\beta
\sigma ^{\prime }}^{\dagger }d_{i\alpha \sigma ^{\prime
}}d_{i\beta \sigma }) \nonumber \\
+J^{\prime }\sum_{\alpha \neq \beta
}d_{i\alpha \uparrow }^{\dagger }d_{i\alpha \downarrow }^{\dagger
}d_{i\beta \downarrow }d_{i\beta \uparrow }.
\label{hi}
\eea
where the interaction parameters were calculated from the comparison
between the energy of adding two holes in the same CoO$_6$ cluster
with given symmetry and spin, or in different clusters. 
The eigenvalues of $H_{I}^{i}$ with two holes 
should coincide 
with the corresponding lowest energy levels
for 6 holes in the cluster. The
resulting parameters of the model become $t=0.10$ eV, $D=0.105$~eV,
$U=1.86$ eV, $U'=1.27$ eV, $J=0.35$ eV and $J'=0.17$ eV. To our knowledge, this the
first time that a calculation of the interaction terms is reported in this system.
Note that the values of the $U$ parameters are smaller than those used in previous
calculations \cite{zhou,ishida-liebsch,marianetti}, but are still much larger than the
bandwidth. We have also added a direct hopping between orbitals
of the same symmetry $t'=0.02$ eV that provide the best agreement
with experiments, however our main results are not affected by the
specific chosen value.

Here we solve $H_{\rm eff}$  using the DMFT \cite{rmp}.
The associated quantum impurity
problem is a three-orbital Anderson impurity that is solved using 
the Hirsch-Fye quantum Monte Carlo (QMC) algorithm \cite{roz}. Due to the symmetry of the 
band structure of $H_0$, the DMFT quantum impurity problem and its
corresponding self-consistency constraint (Dyson Equation) are diagonal in
orbital and spin indexes. 
Thus, the resulting local self-energies $\Sigma^{\rm DMFT}_{\alpha,\sigma}$ 
are also diagonal.
In order to obtain the momentum and energy resolved
Green's functions, the local self-energies have to be analytically continued to
the real frequency domain. Thus, we obtained high quality QMC data using over
one million sweeps to reliably perform the continuation by means of a standard maximum
entropy method \cite{gubernatisjarrell}. 
In the calculations presented here, $J'$ and the spin flip terms in
Eq.~(\ref{hi}) were neglected. This
simplification introduces tiny modifications in the 
results \cite{liebsch-ishida}.
Thus, we adopt the interaction parameters $U_{\alpha,\alpha} = U = 1.86$ eV for the intra-orbital 
repulsion, and 
$U_{\alpha, \alpha'}^{\sigma, -\sigma} = U'=1.27$ eV 
and $U_{\alpha, \alpha'}^{\sigma, \sigma} = U'-J= 0.92$ eV, for the inter-orbital
repulsions with opposite or the same spin, respectively.

The predicted band structure is then obtained from the imaginary part of the lattice
Green's functions given by (we study paramagnetic solutions so we drop the spin
index)
$G_{\alpha}(\bf{k},\omega) = [\omega -  \epsilon_{\bf{k},\alpha} - \Sigma^{DMFT}_{\alpha}(\omega)]^{-1}$
and the Fermi surface is mapped out from the $\omega=0$ crossings of the interacting bands.

We focus our study on 
the cases that are experimentally most relevant, i.e., for doping
$x$=0.3, 0.5 and 0.7,
which range from stronger to weaker correlations. 
For reasons of space, the data displayed in the figures are for $x$=0.3 and 0.7.
Due to their high computational cost, the lowest temperature that we study is $T = 360$~K.
Comparison with calculations at higher temperatures ($\sim 720$~K) indicates that we have 
indeed achieved the low $T$ limit. In addition, as will be shown latter, the width of the 
quasiparticle band at the Fermi energy, 
which is the smallest energy scale in the electronic structure, is much larger than the temperature
of the calculation.

\begin{figure}[ht]
	\includegraphics[width=\linewidth]{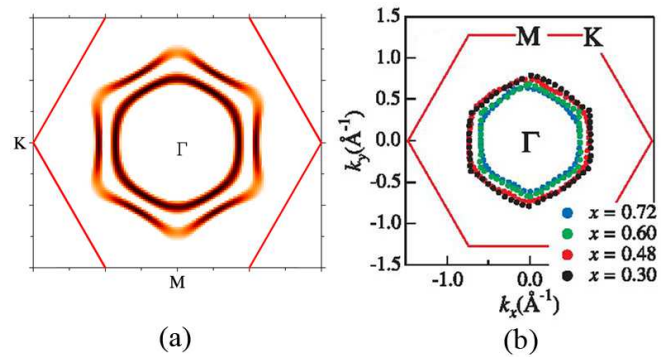}
  \caption{(Color online) (a) Calculated FS for dopings $x=0.3$ and $x=0.7$ 
(inner (hole) edge of larger and smaller hexagon respectively). 
(b) ARPES measurements from~\cite{arpes1}.}
  \label{fig1}
\end{figure}

In Fig.~\ref{fig1} we show our results for the evolution of the Fermi surface as function
of increasing doping along with the respective experimental ARPES data.
We observe good agreement in the shape and size of the FS at
all doping levels. Significantly, the hole pockets are absent in our results.

The experimental FS 
for $x=0.3$ is somewhat more rounded than the theoretical one. This may
partialy be due to the relatively large thermal broadening in the calculation, but may
also be due to lack of hopping terms at longer distances, beyond those included in $H_0$.

\begin{figure}[ht]
	\includegraphics[width=\linewidth]{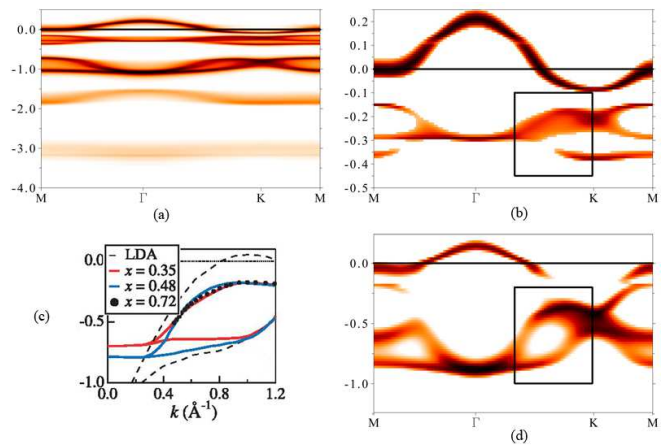}
  \caption{(Color online) (a)  Full band structure for $x=0.3$. The $x=0.3$ 
and $x=0.7$ cases are detailed in (b) and (d) respectively, where the structures in the 
dispersion corresponding to the sinking pockets are highlighted by a box. (c) 
 sinking pockets for various dopings, with comparison to LDA~\cite{arpes1}.}
  \label{fig2}
\end{figure}

The details of the band structure are shown in Fig.~\ref{fig2}. 
We observe that the data reveal several contributions that can be
associated to either coherent (i.e. quasiparticle like) or incoherent (i.e. Hubbard) bands.
The incoherent bands are characterized by dispersive structures similar 
to those of the 
``non-interacting" Hamiltonian $H_0$ (though usually less defined due to 
shorter lifetimes) that appear far from the Fermi energy. These large energy shifts
are of course due to the local interactions of $H_{\rm eff}$ [Eq. (\ref{hi})]. In the top left panel of Fig.~\ref{fig2} we
show 
the full band-structure for the strongly correlated case $x$ = 0.3.
There, one can observe several incoherent bands that appear shifted down in energy, at $\sim$ -1, -1.75 and -3 eV, with
the first one carring a large part of the spectral intensity. Their shapes reveal their dominant orbital content
and their energy shifts can be understood from the values of the inter- and intra-orbital Coulomb repulsions. 
At higher dopings,
the correlation effects decrease and these
incoherent bands rapidly loose spectral intensity.  
On the other hand, the coherent bands are near the Fermi energy and their band structure is
somewhat narrowed with respect to that of $H_0$ [Eq. (\ref{h0})] due to the 
effect of $H_I$ [Eq. (\ref{hi})], indicating the enhancement of the effective mass. 

The top right panel of the figure shows details of the band structure at $x$=0.3 
and, for comparison with the less correlated case, the lower right panel shows 
similar data for $x$=0.7.
Interestingly, these results reveal a novel insight on the nature of the
"sinking pockets", whose experimental data we reproduced in the lower left panel.  
We find that while the sinking pockets are present at both, low and high
dopings (they are indicated by boxes in the respective panels), their physical origin is qualitatively different.
At higher $x$, correlations are low and the band structure does not differ much from the
non-interacting case. Thus, the sinking pocket in this case can be simply associated to the top of the
band with mostly $e'_g$ character [indicated by a box in panel (d)]. 
In contrast, at $x=0.3$, in the strong correlation case, as we
discussed before, the band structure is 
dramatically modified and that interpretation is no longer possible. In fact, the strongest contribution
to the $e'_g$ band is shifted down in energy by about 1 eV. This shift is due to the inter-orbital
Hubbard repulsion, and can be more easily understood in a hole picture. As there is about one hole in the
$a_{1g}$ band, putting a second hole costs $U_{a1g, e'g} \sim 0.92 - 1.27$ eV if the hole goes into
the $e'_g$ band [or $U_{a1g, a1g} \sim 1.86$ eV into the $a_{1g}$ band 
[see panel (a)].
However, in the ground-state there is also a non-negligible amplitude for a configuration with no holes in 
the $a_{1g}$ band, thus one may create a hole in the original $e'_g$ band with no extra Coulomb energy cost.
Such a state would have a reduced spectral intensity but a similar dispersion as that of the 
non-interacting $e'_g$ band, thus accounting for the sinking pocket. 
This is confirmed by our calculations on the weight of the 
states with different symmetries in the coherent bands (not shown).
A clear signature of this sinking pocket state is indicated by a box in our numerical 
data of panel (b).
We note that our results show a lower energy edge at $\sim -0.2$ eV in good
agreement with all available ARPES data \cite{arpes1,arpes2}. Nevertheless, the experimental situation
is less clear for the determination of the dispersive shape of the sinking pockets at higher binding energies.
With regard to the influence of possible Na ordering, 
recent ARPES experiments \cite{arpes2} conclude that 
that feature does not affect significantly the position of the sinking pockets in the unfolded Brillouin zone.
 
In conclusion, motivated by the apparent failure of LDA band structure to provide a sound starting point
for the calculation of strong correlation effects in the cobaltates, we derive an effective 
low-energy Hamiltonian that includes the strength of the local Coulomb repulsive interactions.
The effective model is obtained from finite cluster and quantum chemistry calculations, with essentially no
adjustable parameters.  The effective Hamiltonian is treated with DMFT to compute the effects of 
correlations at different doping levels. 
We find that the evolution of the Fermi surface is in good agreement with the experimental data. Importantly, 
the LDA-predicted hole pockets, that are not seen in the ARPES data, are also not present in our results.
One difference with respect to LDA calculations is how the effect of correlations separates the bands
as discussed above. However, for realistic $U$ this feature alone is not enough to destroy the pockets
and the value of $D$, usually underestimated by LDA, was also shown to play a key role. 
We obtained the detailed interacting electronic structure that reveals sinking pockets at all dopings.
Significantly, their origin is qualitatively different in the high and low doping cases. 

We thank V. Vildosola for help with the implementation of the QMC code, and V. Brouet for discussions on ARPES data. 
This investigation was sponsored by PIP 5254 of CONICET and PICT 2006/483 of the ANPCyT, and by
the ECOS-Sud program. 
AAA is partially supported by CONICET.


\begin{thebibliography}{99}
\bibitem{cdf}M. S. Hybertsen {\it et al.}, Phys. Rev. B {\bf 41}, 11 068
(1990); J. B. Grant and A. K. McMahan, Phys. Rev. Lett. {\bf 66},
488 (1991).

\bibitem{zr} F.C. Zhang and T.M. Rice, Phys. Rev. B {\bf 37}, 3757 (1988).

\bibitem{hub} L. F. Feiner, J. H. Jefferson, and R. Raimondi, Phys. Rev. B {\bf 53},
8751 (1996); M. E. Simon, A. A. Aligia, and E. R. Gagliano, Phys.
Rev. B {\bf 56}, 5637 (1997); references therein.


\bibitem{pt}  A. A. Aligia, M. E. Simon, and C. D. Batista, Phys. Rev. B
\textbf{49}, 13061 (1994).

\bibitem{oles} J. Bala, A. M. Ole\'s, and J. Zaanen, Phys. Rev. Lett. {\bf 72},
2600 (1994).

\bibitem{epl} C. D. Batista, A. A. Aligia, and J. Eroles, Europhys. Lett. {\bf 43}, 71 (1998).


\bibitem{feiner} P. Horsch {\it et al.}, Phys. Rev. Lett. {\bf 101},
167205 (2008); references therein.

\bibitem{arpes1} H. Yang {\it et al.}, Phys. Rev. Lett. {\bf 95}, 146401 (2005).

\bibitem{arpes2} D. Qian {\it et al.} Phys. Rev. Lett. {\bf 97}, 186405 (2006).

\bibitem{lda} D. Singh, Phys. Rev. Lett. {\bf 61}, 13397 (2000) and Phys. Rev. B {\bf 68}, R020503 (2003); P. Zhang {\it et al.}, Phys. Rev. Lett. {\bf 93}, 236402 (2004); K. Lee, J. Kunes, and W. Pickett, Phys. Rev. B {\bf 70}, 045104 (2004).

\bibitem{zhou} S. Zhou {\it et al.}, Phys. Rev. Lett. {\bf 94}, 206401 (2005).

\bibitem{ishida-liebsch} H. Ishida, M. Johannes, and A. Liebsch, Phys. Rev. Lett. {\bf 94}, 196401 (2005).

\bibitem{marianetti} C. Marianetti, K. Haule, and O. Parcollet, Phys. Rev. Lett. {\bf 99}, 246404 (2007).

\bibitem{liebsch-ishida} A. Liebsch and H. Ishida, Eur. Phys. J. B {\bf 61}, 405 (2008). 

\bibitem{vol} J. Kunes {\it et al.}, Phys. Rev. Lett. {\bf 99}, 156404 (2007).

\bibitem{gut} G-T. Wang, X. Dai, and Z. Fang, Phys. Rev. Lett. {\bf 101}, 066403 (2008).

\bibitem{bourgeois} A. Bourgeois {\it et al.}, Phys. Rev. B {\bf 75}, 174518 (2007).

\bibitem{kroll} T. Kroll, A. A. Aligia, and G. Sawatzky, Phys. Rev. B {\bf 74}, 115124 (2006).

\bibitem{wu} W. B. Wu {\it et al.}, Phys. Rev. Lett. {\bf 94}, 146402
(2005).


\bibitem{koshi} W. Koshibae and S. Maekawa, Phys. Rev. Lett. {\bf 91}, 257003 (2003).

\bibitem{ll} S. Landron and M.B. Lepetit, Phys. Rev. B {\bf 74}, 184507 (2006) and 
Phys. Rev. B {\bf 77}, 125106 (2008).

\bibitem{rmp} A. Georges {\it et al.}, Rev. Mod. Phys. {\bf 68}, 13 (1996).

\bibitem{roz} M.J. Rozenberg, Phys. Rev. B {\bf 55}, R4855 (1997).

\bibitem{gubernatisjarrell} J. Gubernatis {\it et al.}, Phys. Rev. B {\bf 44}, 
6011 (1991) and Phys. Reports {\bf 269}, 133 (1996).


\end{thebibliography}
\end{document}